\documentstyle[epsf,aps]{revtex}
\renewcommand{\thefootnote}{\fnsymbol{footnote}}
\begin{document}
\begin{flushright}
BNL-NT-99/6 \\ 
TAUP-2612-99
\end{flushright}
\vspace*{1cm} 
\setcounter{footnote}{1}
\begin{center}
  {\Large\bf Diffractive Dissociation Including Multiple Pomeron
  Exchanges \\ ~~ \\ in High Parton Density QCD} \\[1cm] Yuri V.\
  Kovchegov $^1$ and Eugene Levin $^{1,2}$ \\ ~~ \\ {\it $^1$ Physics
  Department, Brookhaven National Laboratory \\ Upton, NY 11973, USA }
  \\ ~~ \\ {\it $^2$ School of Physics and Astronomy, Tel Aviv
  University \\ Tel Aviv, 69978, Israel }
  \renewcommand{\thefootnote}{\fnsymbol{footnote}}\setcounter{footnote}{0}
  \footnote{Permanent address} \\ ~~ \\ ~~ \\
\end{center}
\begin{abstract} 
 We derive an evolution equation describing the high energy behavior
 of the cross section for the single diffractive dissociation in deep
 inelastic scattering on a hadron or a nucleus.  The evolution
 equation resums multiple BFKL pomeron exchanges contributing to the
 cross section of the events with large rapidity gaps. Analyzing the
 properties of an approximate solution of the proposed equation we
 point out that at very high energies there is a possibility that for
 a fixed center of mass energy the cross section will reach a local
 maximum at a certain intermediate size of the rapidity gap, or,
 equivalently, at some non-zero value of the invariant mass of the
 produced particles.
\end{abstract}

\newcommand{\stackeven}[2]{{{}_{\displaystyle{#1}}\atop\displaystyle{#2}}}
\newcommand{\lsim}{\stackeven{<}{\sim}}
\newcommand{\gsim}{\stackeven{>}{\sim}}
\newcommand{\be}{\begin{eqnarray}}
\newcommand{\dlq}{\lq\lq}
\newcommand{\ee}{\end{eqnarray}}

\section{Introduction}

\begin{figure}
\begin{center}
\epsfxsize=10cm
\leavevmode
\hbox{ \epsffile{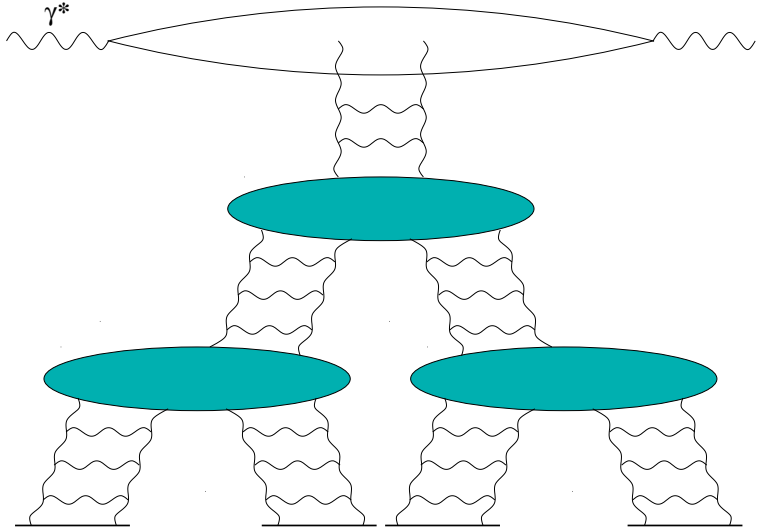}}
\end{center}
\caption{An example of the pomeron ``fan'' diagram, which were considered in 
deriving (\ref{eqN}).}
\label{fan}
\end{figure}

Some time ago an evolution equation was derived in the framework of
Mueller's dipole model \cite{Mueller1,Mueller2,Mueller3,MZ} which
resums multiple Balitsky, Fadin, Kuraev and Lipatov (BFKL)
\cite{EAK,Yay} pomeron exchanges for deep inelastic scattering (DIS)
in the leading $\ln 1/x$ approximation \cite{me}. In terms of the
usual Feynman diagram formulation the equation in \cite{me} sums up
the so-called pomeron ``fan'' diagrams contributing to the total cross
section of a quark--antiquark pair on a hadron or nucleus (see
Fig. \ref{fan}), and turns out to be a generalization of the Gribov,
Levin and Ryskin (GLR) equation \cite{GLR}. Similar nonlinear equation
for multiple pomeron exchanges was also obtained by Balitsky
\cite{Bali} in the effective lagrangian approach.

The physical picture of DIS presented in \cite{me} is the following:
virtual photon splits into a quark--antiquark pair, which, by the time
it reaches the target develops a cascade of dipoles, each of which
independently interacts with the target. All the QCD evolution was
included in the developed cascade of the dipoles in the large $N_c$
limit, similar to \cite{Mueller1,Mueller3}.

The equation in \cite{me} was written for the elastic amplitude of the
scattering of the $q \overline{q}$ pair of transverse size ${\bf x}$
located at impact parameter ${\bf b}$ on a hadron or nucleus, $N_0
({\bf x},{\bf b}, Y)$, where $Y \sim \ln 1/x$ is the rapidity variable
of the slowest particle in the $q \overline{q}$ pair. The target's
structure function $F_2$, as well as the total DIS cross section,
could be easily expressed in terms of $N_0 ({\bf x},{\bf b}, Y )$:
\begin{equation}\label{f2}
  F_2 (x, Q^2) = \frac{Q^2}{4 \pi^2 \alpha_{EM}} \int \frac{d^2 {\bf
  x}_{01} d z }{4 \pi} \, \Phi ({\bf x}_{01},z) \ d^2 b \ 2 \ N_0
  ({\bf x}_{01},{\bf b} , Y) ,
\end{equation}
where the quark and antiquark are located at transverse coordinates
${\bf x}_0$ and ${\bf x}_1$ correspondingly and ${\bf x}_{01} = {\bf
x}_1 - {\bf x}_0$. $Q^2$ is the virtuality of the photon, $z$ is the
fraction of the pair's momentum carried by a quark. $\Phi ({\bf
x}_{01},z)$ is the probability of virtual photon fluctuating into a $q
\overline{q}$ pair, which we will refer to as virtual photon's wave
function. The exact expression for $\Phi ({\bf x}_{01},z)$ is well
known and is given in \cite{me} as well as in a number of other
references.

 In \cite{me} the following equation was written for $N_0 ({\bf
 x},{\bf b}, Y )$:
\begin{eqnarray*}
  N_0 ({\bf x}_{01},{\bf b}, Y) = \gamma ({\bf x}_{01},{\bf b}) \,
  \exp \left[ - \frac{4 \alpha C_F}{\pi} \ln \left(
  \frac{x_{01}}{\rho} \right) Y \right] + \frac{\alpha C_F}{\pi^2}
  \int_0^Y d y \, \exp \left[ - \frac{4 \alpha C_F}{\pi} \ln \left(
  \frac{x_{01}}{\rho} \right) (Y - y) \right]
\end{eqnarray*}  
\begin{eqnarray}\label{eqN}
  \times \int_\rho d^2 x_2 \frac{x_{01}^2}{x_{02}^2 x_{12}^2} \, [ 2
  \, N_0 ({\bf x}_{02},{\bf b} + \frac{1}{2} {\bf x}_{12}, y) - N_0
  ({\bf x}_{02},{\bf b} + \frac{1}{2} {\bf x}_{12}, y) \, N_0 ({\bf
  x}_{12},{\bf b} + \frac{1}{2} {\bf x}_{02}, y) ] ,
\end{eqnarray}
where the initial condition $\gamma ({\bf x}_{01},{\bf b})$ was given
by the Glauber--Mueller interaction formula of the quark--antiquark
pair with the nucleus
\begin{equation}\label{gla}
   \gamma ({\bf x}_{01},{\bf b}_0) = 1 - \exp \left[ - \frac{\alpha
   \pi^2}{2 N_c S_\perp} {\bf x}_{01}^2 A xG ( x ,1/{\bf x}_{01}^2 )
   \right].
\end{equation}
Here, for a cylindrical nucleus, $S_\perp = \pi R^2$ is the transverse
area of the hadron or nucleus, $A$ is the atomic number of the
nucleus, and $xG$ is the gluon distribution in a nucleon in the
nucleus, which was taken at the two gluon (lowest in $\alpha$) level
\cite{Mueller4}. In Eq. (\ref{eqN}) $\rho$ is an ultraviolet
regulator, which never appears in $N_0 ({\bf x}_{01},{\bf b}, Y)$.

There have been several attempts to solve Eq. (\ref{eqN}) analytically
\cite{me2,LT}. Unfortunately it turned out to be very hard to construct 
an analytical solution describing simultaneously the behavior of $N_0
({\bf x}_{01},{\bf b}, Y)$ both outside and inside of the saturation
region. Instead Eq. (\ref{eqN}) was solved separately outside [$k >
Q_s (Y)$] and inside [$k < Q_s (Y)$] the saturation region
\cite{me2,LT}, where approximately \cite{me2}
\begin{equation}
Q_s (Y) \sim \Lambda \, \alpha^2 \, A^{1/3} \, \frac{\exp [ (\alpha_P
- 1)Y] }{\sqrt{14 \alpha N_c \zeta (3) Y}},
\end{equation}
with $\alpha_P - 1 = \frac{4 \alpha C_F}{\pi} \, \ln 2$. In
\cite{me2,LT} it was concluded that for not very large energies,
corresponding to rapidities of the order of $Y \sim 1/\alpha$ (or,
equivalently, for $k > Q_s (Y)$ ), the solution of Eq. (\ref{eqN})
behaves with energy like a single BFKL pomeron exchange, with multiple
pomeron corrections setting in as energy increases and slowing down
the growth of $N_0 ({\bf x}_{01},{\bf b}, Y)$. For very high energies
($Y \ge \frac{1}{\alpha_P - 1} \ln \frac{1}{\alpha^2}$ or $k < Q_s
(Y)$) the solution of Eq. (\ref{eqN}) saturates to a constant, namely
$N_0 ({\bf x}_{01},{\bf b}, Y) = 1$, which corresponds to the
blackness of the total cross section of the quark--antiquark pair on
the nucleus \cite{me2,LT}. Thus, as one can see at least the
qualitative features of $N_0 ({\bf x}_\perp,{\bf b}, Y)$ as a function
of $Y$ and $x_\perp$ are known. For a better quantitative
understanding of the behavior of $N_0 ({\bf x}_\perp,{\bf b}, Y)$ one
should, probably, solve Eq. (\ref{eqN}) numerically, which would be a
very interesting and useful project to perform.

\begin{figure}
\begin{center}
\epsfxsize=8cm
\leavevmode
\hbox{ \epsffile{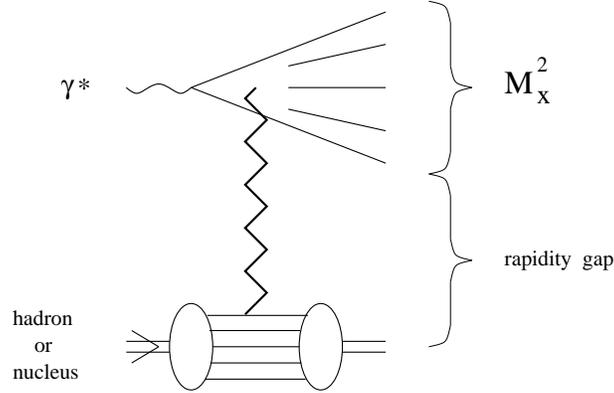}}
\end{center}
\caption{Single diffractive dissociation process considered in the paper.
 The interaction between the target and virtual photon is represented 
schematically by an exchange of a color singlet object.}
\label{rapgap}
\end{figure}

In \cite{me} the object of interest was the total inclusive DIS cross
section, where no restrictions are imposed on the final state of the
process.  In this paper we are going to study the cross section of the
single diffractive dissociation. The physical picture of the process
we are going to consider is the following: in DIS the virtual photon
interacts with the hadron or nucleus breaking up into hadrons and jets
in the final state. At the same time the target hadron (nucleus)
remains intact. The particles produced as a result of hadron's breakup
do not fill the whole rapidity interval, leaving a rapidity gap
between the target and the ``slowest'' produced particle. The process
is depicted in Fig. \ref{rapgap}. In diffractive dissociation one
imposes a restriction on the final state --- the existence of a
rapidity gap.

\begin{figure}
\begin{center}
\epsfxsize=8cm
\leavevmode
\hbox{ \epsffile{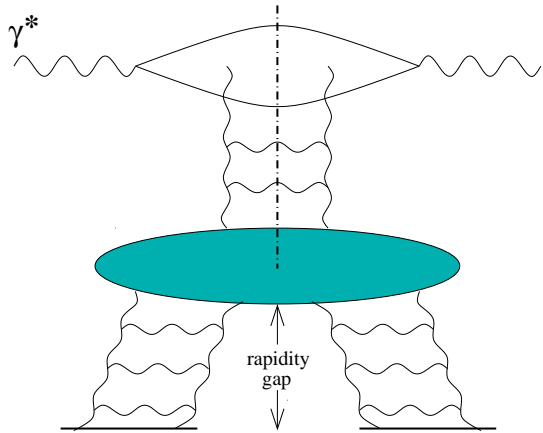}}
\end{center}
\caption{Traditional description of single diffractive dissociation. 
Dash-dotted line represents the final state.}
\label{3pom}
\end{figure}

Below we are going to employ the techniques of Mueller's dipole model
\cite{Mueller1,Mueller2,Mueller3,MZ}, similarly to the way they were applied in
\cite{me}, to construct a cross section of the single diffractive 
dissociation in DIS which would include the effects of multiple
pomeron exchanges. Analogous to \cite{me,me2} we will neglect the
diagrams with pomeron loops, i.e. , the diagrams where a pomeron
splits into two pomerons and then the two pomerons merge into one
pomeron. As was argued in \cite{me,me2,shw,AGL} these diagrams are
suppressed compared to the pomeron fan diagrams of Fig. \ref{fan} in
DIS on a large nucleus by factors of $A^{1/3}$. Thus one can safely
neglect them up to rapidities of the order of $Y \gsim
\frac{1}{\alpha_P -1} \ln \frac{1}{\alpha^2 A^{1/3}}$. For the case of 
DIS on a proton, strictly speaking there is no similar argument
allowing one to neglect pomeron loop diagrams. Therefore, whereas the
fan diagram evolution dominates in DIS on a nucleus, it could only be
considered as a model for DIS on a hadron.

In the traditional description of the single diffractive dissociation
one usually considers triple pomeron vertex \cite{Mueller2,3p,agk,wm},
where the pomeron above the vertex is cut, and the two pomerons below
the vertex are on different sides of the cut, thus producing a
rapidity gap, as shown in Fig. \ref{3pom}. That way the particles are
produced by the cut pomeron in the rapidity interval adjacent to the
virtual photon's fragmentation region and no particles are produced
with rapidities close to the final state of the target due to uncut
pomerons. In this paper we want to enhance this picture by including
multiple pomeron exchange diagrams. We would like to understand the
behavior of the diffractive dissociation at very high energies, which
may include some effects of saturation of hadronic or nuclear
structure functions.

\begin{figure}
\begin{center}
\epsfxsize=10cm
\leavevmode
\hbox{ \epsffile{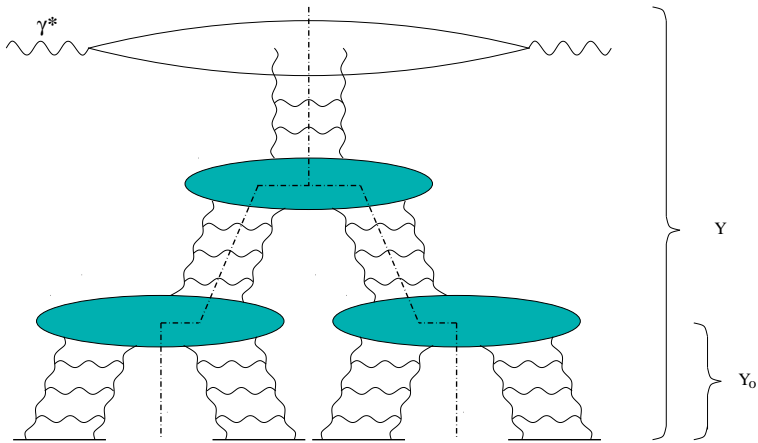}}
\end{center}
\caption{A diagram contributing to the diffractive dissociation with 
 rapidity gap $Y_0$ as considered in the text.}
\label{fandd}
\end{figure}

An example of a graph in the usual Feynman diagram language which we
will consider below is given in Fig. \ref{fandd}. Dash-dotted line
corresponds to the final state, i.e., to the cut. As one can see in
Fig. \ref{fandd} some of the pomerons are cut, some remain uncut. In
the region of rapidity where pomerons are cut we have particles being
produced. In the notation of Fig. \ref{fandd} this corresponds to the
interval in rapidity from $Y - Y_0$ to $Y$. In the region where the
pomerons are uncut (rapidity interval from $0$ to $Y_0$) nothing is
produced, which corresponds to a rapidity gap. That way
Fig. \ref{fandd} demonstrates a generalization of the traditional
picture of Fig. (\ref{3pom}), in which the cut pomeron splits into two
pomerons, which later branch into two uncut pomerons each. Our goal in
this paper is to resum all the diagrams where the cut pomeron can
split into any number of cut pomerons via fan diagrams, and the cut
pomerons in turn split into uncut pomerons, which can also branch into
any number of uncut pomerons interacting with the target below.

In Sect. II we employ the dipole model formalism
\cite{Mueller1,Mueller2,Mueller3,MZ} to write down an equation for the 
cross section of the single diffractive dissociation of a
quark--antiquark pair on a hadron or nucleus with the rapidity gap
bigger or equal to $Y_0$.  The task is not very straightforward, since
the dipole model provides us with the dipole wave function of the $q
\overline{q}$ pair by the time it hits the target. In the rest frame
of the target the typical time scale associated with the interaction
is negligibly small compared to the typical lifetime of the QCD
evolved wavefunction of the pair. Thus one could say that the
interaction is instantaneous and happens at the time $t=0$
\cite{km}. However, after the interaction, the $q
\overline{q}$ state with many gluons in it can undergo several 
transformation until all the partons in it reach the mass shell at
$t=\infty$. Dipole model does not provide us with any information
about these final state interactions. Nevertheless one can gain
control over the final state through the unitarity condition and
cancelation of certain classes of final state interactions, as will be
shown below. As a result we obtain a nonlinear evolution equation for
the diffractive cross section, shown in formula (\ref{eqND}).

We show how Eq. (\ref{eqND}) can be re-derived by applying the AGK
cutting rules to the pomeron fan diagrams in Sect. III. Since the cut
or uncut pomerons precisely specify the final state in the traditional
language, we, therefore, proved the consistency of our treatment of
the final states in the dipole model.

Even though the exact analytical solution of that equation seems to be
very complicated in Sect. IV we analyze the properties of the
solution. As a toy model we consider a simplified version of the
equation with the kernel independent of transverse coordinates. We
show that the simplified equation's solution exhibits a remarkable
property at very high energies. We plot the cross section of
diffractive dissociation events with rapidity gap $Y_0$ at fixed large
center of mass energy as a function of the size of the rapidity gap
$Y_0$. This is equivalent to plotting the cross section as a function
of the invariant mass of the produced particles $M_X^2$, since, as
could be easily seen from Fig. \ref{rapgap}, $Y_0 = Y - \ln
M_X^2/Q^2$. At not very large $Y_0$ (large $M_X^2$) the cross section
increases with the increase of rapidity gap, which agrees with the
result of the traditional triple pomeron description of the
diffractive dissociation of \cite{Mueller2,3p}. However, as $Y_0$ gets
very high comparable to the values of rapidity at saturation, which
corresponds to a very small produced mass $M_X^2$, the cross section
reaches a maximum and starts decreasing. That way if one would be able
to measure the single diffractive cross section for a range of
different invariant masses of the produced particles, one should be
able to observe the maximum and turnover of the cross section for
small masses if the energy is high enough for the saturation effects
to become important. If the exact, probably numerical, solution of the
equation confirms our conclusion, which was reached by approximate
methods, then the single diffractive cross section is a new and
independent observable which can be used to test whether the
saturation region has been reached in DIS experiments.

 The summary of our results is presented in Sect. V.

\section{Evolution equation for diffractive cross section}

In this section we will write down an evolution equation resumming
multiple pomeron exchanges for the cross section of single diffractive
dissociation in DIS on a hadron or nucleus (see Figs. \ref{rapgap} and
\ref{fandd}). As was mentioned in the introduction the task is
complicated by the fact that the techniques of the dipole model
\cite{Mueller1,Mueller2,Mueller3,MZ} which we are going to employ
provide us with the state of the system at time $t=0$, without putting
any restrictions on its subsequent time evolution until
$t=\infty$. The dipole cascade is developed by the time of interaction
and interacts with the nucleus at $t=0$ for any process. It is the
subsequent time evolution of the cascade that determines whether the
event is going to have a rapidity gap, which would correspond to some
of the dipoles recombining, or not. This evolution is not included in
the dipole model \cite{Mueller1,Mueller2,Mueller3,MZ}.  However we
will be able to constrain that time evolution after the interaction
took place and to construct the diffractive cross section using the
following two observations.

\begin{figure}
\begin{center}
\epsfxsize=10cm
\leavevmode
\hbox{ \epsffile{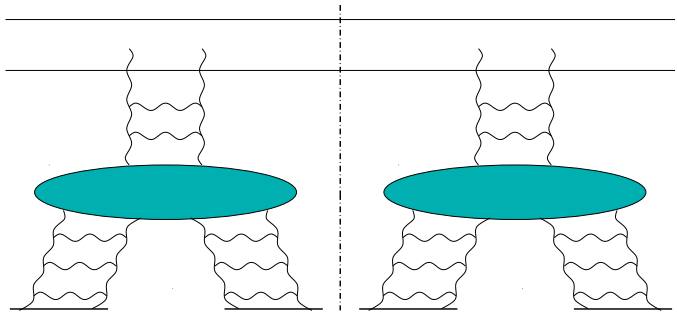}}
\end{center}
\caption{An example of elastic scattering of the quark--antiquark pair 
on the nucleus as pictured in the text.}
\label{el}
\end{figure}

The first observation is that if one knows the total cross section for
a the scattering of some probe on a target one can easily construct
the elastic scattering cross section out of it using unitarity
constraints. Really, if the total cross section is \cite{mL,mu,Fro}
\begin{mathletters}
\begin{eqnarray}\label{stot}
\sigma_{tot} = 2 \int d^2 b \, \, [1 - S(b)] 
\end{eqnarray}
with $S(b)$ the $S$--matrix of the scattering process at a fixed
impact parameter $b$, then the elastic cross section is
\cite{mL,mu,Fro}
\begin{eqnarray}\label{sel}
\sigma_{el} =  \int d^2 b \, \, [1 - S(b)]^2 .
\end{eqnarray}
\end{mathletters}
Using Eqs. (\ref{stot}),(\ref{sel}) it was shown in \cite{me2,mL} that
the elastic cross section for the scattering of the quark--antiquark
pair generated by a virtual photon in DIS on a nucleus or hadron could
be expressed in terms of the elastic amplitude $N_0 ({\bf x}_{01},{\bf
b}, Y)$, or, equivalently, in terms of the imaginary part of the
forward scattering amplitude $N_0 ({\bf x}_{01},{\bf b}, Y) d^2 b$
(The forward amplitude is purely imaginary.). The result was
\cite{me2,mL}
\be\label{elast}
N^{el}({\bf x}_{01},{\bf b}, Y) = N_0 ({\bf x}_{01},{\bf b}, Y)^2
\ee
and is illustrated in Fig. \ref{el}. The cut in Fig. \ref{el}
corresponds to the final state, when the particles reached the mass
shell and $t=\infty$. Thus in that picture we explicitly impose the
constraint that not only the $q \overline{q}$ system developed a
dipole cascade and the cascade in turn interacted with the target at
$t=0$, but it also recombined back into a $q \overline{q}$ pair by the
time the system reached the final state at
$t=\infty$. Eq. (\ref{elast}) allows us to calculate the elastic cross
section of the scattering of any (not necessarily original) dipole in
the dipole cascade on the nucleus including multiple pomeron
evolution.

\begin{figure}
\begin{center}
\epsfxsize=12cm
\leavevmode
\hbox{ \epsffile{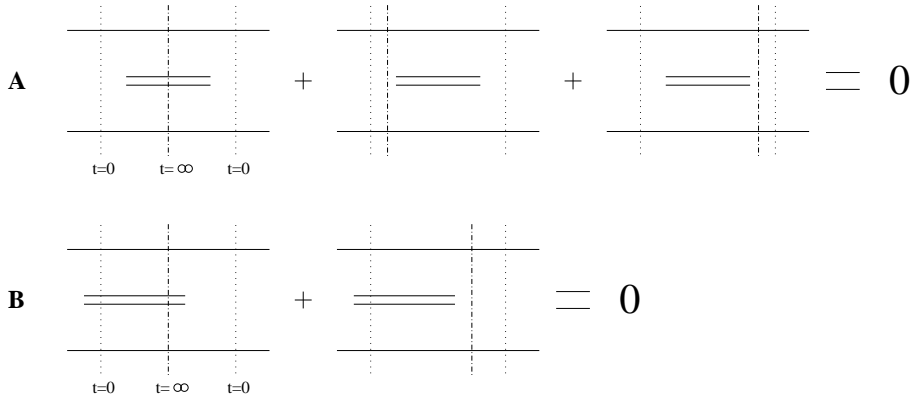}}
\end{center}
\caption{Classes of diagrams with final state interactions which cancel.}
\label{canc}
\end{figure}

The second observation which we have to make before starting to derive
our equation is based on the work of Chen and Mueller \cite{MZ} and
concerns the so-called final state interactions. By that we mean the
interactions such as branching or recombination of partons in the
dipole cascade after $t=0$, i.e., after the interaction with the
target. Chen and Mueller \cite{MZ} succeeded in proving that certain
classes of final state interactions cancel (see Fig. \ref{canc}).

We denote a gluon by the double line in Fig. \ref{canc} corresponding
to the large $N_c$ limit. In each diagram of Fig. \ref{canc} we assume
that the summation over all possible connections of the gluon to the
quark and antiquark lines is performed.  In Fig. \ref{canc}A we
consider the situation when the gluon is emitted in a dipole after the
interaction with the target at $t=0$, which is denoted by dotted
line. The three different positions of the $t=\infty$ final state cut,
denoted by dash-dotted line in Fig. \ref{canc}A correspond to the
cases when the emitted gluon is present in the final state and when
the gluon recombines back in the amplitude or complex conjugate
amplitude leaving the dipole intact by $t=\infty$. As was proved in
\cite{MZ} the sum of the three cuts of Fig. \ref{canc}A is zero.

In Fig. \ref{canc}B we consider the situation when the gluon was
developed in the dipole wave function before the interaction with the
target and is already there by the time $t=0$. Then it can either
remain in the final state $t=\infty$ or it can recombine back, leaving
only the original dipole in which it was produced in the final
state. In Fig. \ref{canc}B we explore the case when the gluon in the
final state becomes a gluon emitted after $t=0$ in the complex
conjugate amplitude (first diagram in Fig. \ref{canc}B). Chen and
Mueller \cite{MZ} showed that this diagram is canceled by the second
diagram in Fig. \ref{canc}B.

Diagrams of Fig. \ref{canc} demonstrate that the contribution of
gluons produced or absorbed after $t=0$ cancels out in calculation of
inclusive quantities such as the total cross section.  We can now
conclude that the total inclusive DIS cross section can be calculated
using just $t=0$ formalism, as was done in \cite{me}. Really, as one
can see from Fig. \ref{canc} all the final state interactions cancel,
effectively leaving the $t=0$ state unchanged by $t=\infty$. This
conclusion is intuitively easy to understand: one just needs to have
some interaction of the projectile with the target to include the
diagram in the total cross section, independent of what happened after
that interaction.

Now we are ready to write down our equation for diffractive cross
section. Let us start by defining the object for which the equation
will be written. We denote by $N^D ({\bf x}_\perp,{\bf b}, Y, Y_0)$
the cross section of single diffractive dissociation of a dipole of
transverse size ${\bf x}_\perp$, rapidity $Y$ and impact parameter
${\bf b}$ on a target hadron or nucleus. The process has a rapidity
gap covering a rapidity interval adjacent to the target (see
Fig. \ref{rapgap},\ref{fandd} ) which could be greater or equal than
$Y_0$. The corresponding diffractive structure function, similarly to
Eq. (\ref{f2}), is
\begin{equation}\label{f2sd}
  F_2^{SD} (x, Q^2, Y_0) = \frac{Q^2}{4 \pi^2 \alpha_{EM}} \int \frac{d^2
  {\bf x}_{01} d z }{4 \pi} \, \Phi ({\bf x}_{01},z) \ d^2 b \ N^D
  ({\bf x}_{01},{\bf b} , Y, Y_0) .
\end{equation}
If one wishes to obtain a cross section of diffractive dissociation of
the dipole with a fixed rapidity gap $Y_0$ one has to differentiate
$N^D$ with respect to $Y_0$, as will be discussed later. When $Y=Y_0$
the rapidity gap fills out the whole rapidity interval, turning the
process of dipole's dissociation into an elastic scattering. Taking
the answer for elastic process from Eq. (\ref{elast}) we obtain an
initial condition for the evolution of $N^D$
\be\label{incon}
N^D ({\bf x}_\perp,{\bf b}, Y=Y_0, Y_0) = N_0^2 ({\bf x}_\perp,{\bf
b}, Y_0).
\ee

Note that Eq. (\ref{incon}) insures that after the interaction with
the target at $t=0$ the system of developed dipoles only recombines
back into the original dipole by $t=\infty$.  All other possible final
state fluctuations have been excluded because we explicitly put $N_0^2
({\bf x}_\perp,{\bf b}, Y_0)$ as the amplitude of the elastic process.

To construct $N^D$ one has to require that nothing is produced in the
$t=\infty$ final state in the rapidity interval from $0$ to $Y_0$,
where the target hadron or nucleus is situated at rapidity $0$. This
does not restrain the rapidity gaps from being greater than
$Y_0$. That way $N^D$ would include all events with rapidity gap
greater or equal to $Y_0$. The condition of the rapidity gap up to
$Y_0$ is satisfied by Eq. (\ref{incon}).

As was mentioned above our physical picture of the diffractive
dissociation event is the following: before hitting the target the
virtual photon develops a dipole cascade, which at the time $t=0$
interacts with the target.  After that some partons in the cascade may
recombine, producing rapidity gaps, some may not. By imposing the
recombination condition in Eq. (\ref{incon}) we made sure that the
dipoles with rapidities $y<Y_0$ will recombine by $t=\infty$,
producing a rapidity gap.  The dipoles with $y>Y_0$ are free to either
recombine into the dipoles off which they were produced or to remain
unchanged till $t=\infty$. As long as $y>Y_0$ there are no
restrictions on the final state like rapidity gaps. Therefore the
dipoles are free to recombine after $t=0$. However, for that situation
the cancelation rules proven by Chen and Mueller \cite{MZ} apply. The
graphs when the dipoles recombine to either increase the size of the
rapidity gap $Y_0$ or to produce extra rapidity gaps at $y>Y_0$ are
included, but as one can see from Fig. \ref{canc}B they cancel. For
the asymmetric case when the dipole in consideration interacts with
the target in the amplitude and then recombines back into the dipole
off which it was produced which has no interaction with the target in
the complex conjugate amplitude we employ a different cancelation
mechanism: the contributions when the dipole recombines in the complex
conjugate amplitude at the times $t>0$ and $t<0$ cancel each other,
leaving us with the diagram of similar to the second graph of
Fig. \ref{canc}B, which could be incorporated in the elastic
interaction of the original dipole with the target $N_0 ({\bf
x}_\perp,{\bf b}, y)$. That way the asymmetric graphs may include the
dipoles of any rapidities $y>Y_0$ interacting elastically with the
target via Eq. (\ref{incon}). For the ``symmetric'' case when the
dipole at $y>Y_0$ interacts elastically with the target in the
amplitude and then recombines into the original dipole which interacts
with the target elastically in the complex conjugate amplitude we
applied the cancelation of Fig. \ref{canc}B. Thus the symmetric case
reduces to the situation when the same dipole interacts elastically in
the amplitude and complex conjugate amplitude at $y=Y_0$ and is put
only in the initial condition of our equation given by $N_0^2 ({\bf
x}_\perp,{\bf b}, Y_0)$.

Dipole splitting is a little more complicated. The softest dipoles in
the cascade interact the way it is described in Eq. (\ref{incon}),
which fixes the final state for them. Now let us consider a dipole
which is not one of these soft dipoles and, therefore, does not
interact with the target. The dipole can split into two dipoles either
before or after $t=0$. This event is allowed only when the produced
real gluon has the rapidity greater or equal than $Y_0$, so that it
would not destroy the rapidity gap. In that case the real gluon
emitted before $t=0$ is canceled by a virtual correction in the same
dipole before $t=0$. The real gluon emitted after $t=0$ is canceled
by the virtual term after $t=0$, similarly to
Fig. \ref{canc}A. Finally, this non-interacting dipole can develop a
virtual corrections with the gluon being softer than $Y_0$,
i.e. having smaller rapidity than the rapidity of the gap. These gluons
could not be canceled by the real contributions, since those are
excluded by the condition of the gap's existence. However, since the
virtual corrections could happen either before or after $t=0$, these
two combinations come with different signs.  The absolute values of
both $t>0$ and $t<0$ contributions in that case are the same and,
therefore, they cancel each other. Thus we have shown that the final
state dipole splitting is canceled together with the virtual
corrections in the non-interacting dipoles.

\begin{figure}
\begin{center}
\epsfxsize=12cm
\leavevmode
\hbox{ \epsffile{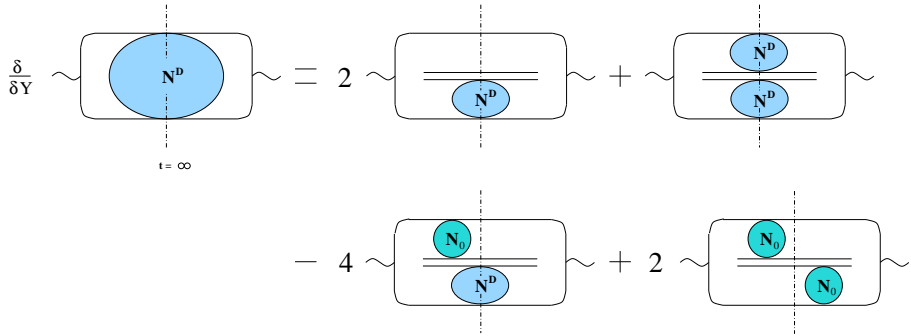}}
\end{center}
\caption{The equation for the cross section of diffractive dissociation.}
\label{eqn}
\end{figure}

The above discussion is summarized in Fig. \ref{eqn}. The change of
the cross section of diffractive dissociation $N^D$ in one step of the
rapidity evolution could be generated by several terms on the right
hand side of the equation in Fig. \ref{eqn}. The first terms
corresponds to the usual single BFKL pomeron evolution, while the rest
of the terms correspond to the triple pomeron vertex, similarly to the
equation written in \cite{Mueller1,me}. Effectively Fig. \ref{eqn}
states that in one step of evolution the color dipole splits into two
dipoles. The subsequent evolution can happen in one of the dipoles,
which corresponds to the first term on the right hand side of
Fig. \ref{eqn}. It could also happen that we continue the evolution
leading to the cross section of diffractive dissociation in both
dipoles (the second terms in Fig. \ref{eqn}). The third and forth
terms correspond to the asymmetric cases. As was discussed above in
the asymmetric case the dipole can interact elastically with the
target in the amplitude but not in the complex conjugate amplitude and
vice versa at any rapidity $Y > Y_0$. Thus we include the cases when
only one of the produced dipoles interacts asymmetrically with the
target (the third term in Fig. \ref{eqn}) and when both of them do so
(the fourth term in Fig. \ref{eqn}). The coefficients in front of
different terms come from combinatorics. Note that they match the
coefficients of the Abramovskii, Gribov, Kancheli (AGK) \cite{agk}
cutting rules for two pomeron contributions.

Recalling the initial condition of Eq. (\ref{incon}) we can turn the
picture of Fig. \ref{eqn} into a formula. For that we also have to
include the virtual corrections similarly to the way they were
included in \cite{Mueller1}. The resulting equation is
\begin{eqnarray*}
  N^D ({\bf x}_{01},{\bf b}, Y, Y_0) = N_0^2 ({\bf x}_{01},{\bf b},
  Y_0) \, e^{ - \frac{4 \alpha C_F}{\pi} \ln \left(
  \frac{x_{01}}{\rho} \right) (Y - Y_0) } + \frac{\alpha C_F}{\pi^2}
  \int_{Y_0}^Y d y \, e^{ - \frac{4 \alpha C_F}{\pi} \ln \left(
  \frac{x_{01}}{\rho} \right) (Y - y)}
\end{eqnarray*}  
\begin{eqnarray*}
  \times \int_\rho d^2 x_2 \frac{x_{01}^2}{x_{02}^2 x_{12}^2} \, [ 2
  \, N^D ({\bf x}_{02},{\bf b} + \frac{1}{2} {\bf x}_{12}, y, Y_0) +
  N^D ({\bf x}_{02},{\bf b} + \frac{1}{2} {\bf x}_{12}, y, Y_0) \, N^D
  ({\bf x}_{12},{\bf b} + \frac{1}{2} {\bf x}_{02}, y, Y_0) 
\end{eqnarray*}  
\begin{eqnarray}\label{eqND}
 - 4 \, N^D ({\bf x}_{02},{\bf b} + \frac{1}{2} {\bf x}_{12}, y, Y_0)
 \, N_0 ({\bf x}_{12},{\bf b} + \frac{1}{2} {\bf x}_{02}, y) + 2 \,
 N_0 ({\bf x}_{02},{\bf b} + \frac{1}{2} {\bf x}_{12}, y) \, N_0 ({\bf
 x}_{12},{\bf b} + \frac{1}{2} {\bf x}_{02}, y) ] .
\end{eqnarray}
The shifts in the impact parameter dependence of the functions on the
right hand side of Eq. (\ref{eqND}) have occurred because the centers
of mass of the produced dipoles are located at different impact
parameters than the center of mass of the initial dipole. 
 
Eq. (\ref{eqND}) describes the small-$x$ evolution of the cross
section of the single diffractive dissociation for DIS $N^D$ and is
the central result of this paper.

\section{Rederiving evolution equation with AGK cutting rules}

Eq. (\ref{eqND}) can also be derived using the technique employing the
AGK cutting rules \cite{agk}. In this section we will not employ the
dipole model. Instead we will consider pomeron fan diagrams in the
usual perturbation theory, imposing a restriction that there are only
triple pomeron vertices in the theory. This assumption is true only in
the large $N_c$ limit, as was shown in the dipole model. 

AGK cutting rules \cite{agk} apply to the BFKL pomerons in QCD (see
\cite{br} for a detailed discussion). For the BFKL pomeron one can write
\be \label{cutp}
2 \, \mbox{Im} \,  a^{BFKL}_{el} = G_{in}^{BFKL},
\ee
where $a^{BFKL}_{el}$ is the one BFKL pomeron contribution to the
forward scattering amplitude and $G_{in}^{BFKL}$ is the total
inelastic cross section. In writing Eq. (\ref{cutp}) we have neglected
the elastic contribution to the cross section, which is a justified
assumption for the BFKL pomeron exchange \cite{br}. Eq. (\ref{cutp})
tells us that the inelastic cross section in one pomeron exchange
approximation is twice the forward amplitude of the process, since the
latter is purely imaginary.

\begin{figure}
\begin{center}
\epsfxsize=15cm
\leavevmode
\hbox{ \epsffile{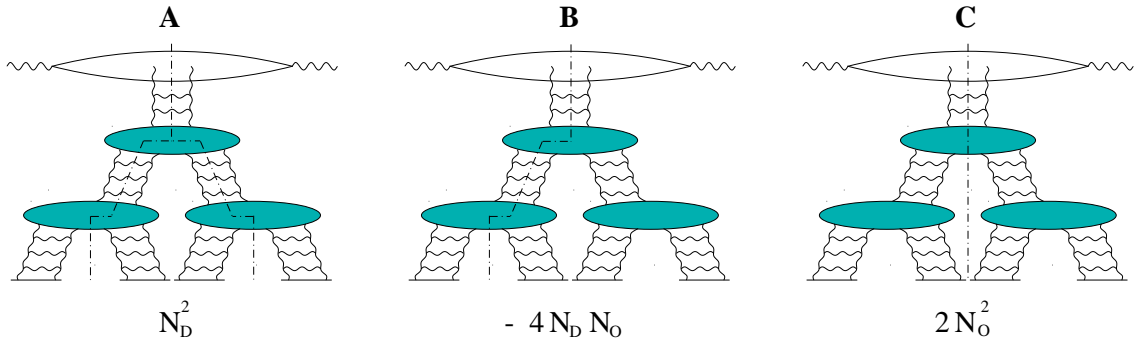}}
\end{center}
\caption{Different pomeron cuts  contributing to the cross section 
of diffractive dissociation which lead to different terms on the right
hand side in Eq. (\ref{eqND}).}
\label{agkfig}
\end{figure}

Different cuts of pomeron fan diagrams corresponding to the terms on
the right hand side of Eq. (\ref{eqND}) are displayed in
Fig. \ref{agkfig}. The first (linear) term on the right hand side of
Eq. (\ref{eqND}) corresponds to just usual linear BFKL evolution with
the factor of $2$ arising from Eq. (\ref{cutp}) and is not shown in
Fig. \ref{agkfig}. The non-linear evolution corresponds to a pomeron
splitting into two via the triple pomeron vertices. Let us now
concentrate on the first (topmost) triple pomeron vertex in the
diagrams of Fig. \ref{agkfig}. This is where one step of non-linear
evolution takes place in our picture. The graph of Fig. \ref{agkfig}A
corresponds to the case when a cut pomeron splits into two cut
pomerons, which later on produce rapidity gaps. That is the cross
section of diffractive dissociation $N_D$ splits into two similar
cross sections. Therefore this term gives rise to $N_D^2$ term in
Eq. (\ref{eqND}). The factor in front of this term results from
multiplying $2$ from the cut pomeron on top of the vertex by the
symmetry factor of $1/2$. The diagram of Fig. \ref{agkfig}B
corresponds to a cut pomeron splitting into another cut pomeron, which
then develops a rapidity gap ($N_D$), and an uncut pomeron,
corresponding to the forward scattering amplitude $N_0$. Since there
are two of such diagrams, multiplying the factor of $2$ of the cut
pomeron by $2$ and putting a minus sign since one of the resulting
pomerons is completely on one side of the cut we will get $- 4 N_D
N_0$, which is the contribution of that diagram in
Eq. (\ref{eqND}). Finally the graph of Fig. \ref{agkfig}C corresponds
to the case when both of the pomerons below the vertex are uncut. Each
uncut pomeron gives a factor of $N_0$. The factor of $2$ arises from
the cut pomeron above [Eq. (\ref{cutp})].

Therefore we were able to reproduce all the terms on the right hand
side of Eq. (\ref{eqND}) from the usual pomeron fan diagrams using the
AGK cutting rules. We note that a similar result was obtained in the
framework of reggeon theory in \cite{pomdif}. The cut (or uncut)
pomeron specifies $t=\infty$ final state very precisely: if the
pomeron is cut it insures that there will be particles produced in the
final state. Uncut pomeron insures that nothing is produced in the
$t=\infty$ final state. Thus in this section we have demonstrated the
consistency of our analysis of the evolution of the dipole state after
the interaction with the target which was employed in the previous
section.

\section{Properties of the solution of the evolution equation}

Eq. (\ref{eqND}) can be rewritten in the differential form. Similar to
what was done in \cite{me} we differentiate both sides of
Eq. (\ref{eqND}) with respect to $Y$ and once again make use of
Eq. (\ref{eqND}) to obtain
\begin{eqnarray*}
  \frac{\partial N^D ({\bf x}_{01},{\bf b}, Y, Y_0)}{\partial Y} =
  \frac{2 \alpha C_F}{\pi^2} \, \int_\rho d^2 x_2 \left[
  \frac{x^2_{01}}{x^2_{02} x^2_{12}} - 2 \pi \delta^2 ({\bf x}_{01}
  -{\bf x}_{02}) \ln \left( \frac{x_{01}}{\rho} \right) \right] N^D
  ({\bf x}_{02},{\bf b} + \frac{1}{2} {\bf x}_{12}, Y, Y_0)
\end{eqnarray*}
\begin{eqnarray*}
  + \frac{\alpha C_F}{\pi^2} \, \int d^2 x_2 \,
  \frac{x^2_{01}}{x^2_{02} x^2_{12}} \, [ N^D ({\bf x}_{02},{\bf b} +
  \frac{1}{2} {\bf x}_{12}, Y, Y_0) \, N^D ({\bf x}_{12},{\bf b} +
  \frac{1}{2} {\bf x}_{02}, Y, Y_0)
\end{eqnarray*}
\begin{eqnarray}\label{diff}
- 4 \, N^D ({\bf x}_{02},{\bf b} + \frac{1}{2} {\bf x}_{12}, Y, Y_0)
\, N_0 ({\bf x}_{12},{\bf b} + \frac{1}{2} {\bf x}_{02}, Y) + 2 \, N_0
({\bf x}_{02},{\bf b} + \frac{1}{2} {\bf x}_{12}, Y) \, N_0 ({\bf
x}_{12},{\bf b} + \frac{1}{2} {\bf x}_{02}, Y) ] ,
\end{eqnarray}
with Eq. (\ref{incon}) being the initial condition for the
differential equation (\ref{diff}).

Let us define the following object
\be\label{fdef}
{\cal F} ({\bf x}_{01},{\bf b}, Y, Y_0) = 2 \, N_0 ({\bf x}_{01},{\bf
b}, Y) - N^D ({\bf x}_{01},{\bf b}, Y, Y_0) \, \theta (Y - Y_0),
\ee
which has the meaning of the cross section of the events with rapidity
gaps less than $Y_0$. The theta-function that multiplies $N^D$ in
Eq. (\ref{fdef}) insures the trivial fact that the rapidity gap can
not be larger than the total rapidity interval. One can see that
Eq. (\ref{diff}) can be rewritten for $Y_0 \le Y$ in terms of ${\cal
F}$ as
\begin{eqnarray*}
  \frac{\partial {\cal F} ({\bf x}_{01},{\bf b}, Y, Y_0)}{\partial Y}
  = \frac{2 \alpha C_F}{\pi^2} \, \int_\rho d^2 x_2 \left[
  \frac{x^2_{01}}{x^2_{02} x^2_{12}} - 2 \pi \delta^2 ({\bf x}_{01}
  -{\bf x}_{02}) \ln \left( \frac{x_{01}}{\rho} \right) \right] {\cal
  F} ({\bf x}_{02},{\bf b} + \frac{1}{2} {\bf x}_{12}, Y, Y_0)
\end{eqnarray*}
\begin{eqnarray}\label{fdiff}
  - \frac{\alpha C_F}{\pi^2} \, \int d^2 x_2 \,
  \frac{x^2_{01}}{x^2_{02} x^2_{12}} \, {\cal F} ({\bf x}_{02},{\bf b}
  + \frac{1}{2} {\bf x}_{12}, Y, Y_0) \, {\cal F} ({\bf x}_{12},{\bf
  b} + \frac{1}{2} {\bf x}_{02}, Y, Y_0),
\end{eqnarray}
with the initial condition 
\be\label{inf}
{\cal F} ({\bf x}_{01},{\bf b}, Y=Y_0, Y_0) = 2 \, N_0 ({\bf
x}_{01},{\bf b}, Y_0) - N_0^2 ({\bf x}_{01},{\bf b}, Y_0).
\ee
If one differentiates Eq. (\ref{eqN}) with respect to $Y$ one would
obtain the same differential equation as Eq. (\ref{fdiff}) for $N_0$
\begin{eqnarray*}
  \frac{\partial N_0 ({\bf x}_{01},{\bf b}, Y)}{\partial Y} = \frac{2
  \alpha C_F}{\pi^2} \, \int_\rho d^2 x_2 \left[
  \frac{x^2_{01}}{x^2_{02} x^2_{12}} - 2 \pi \delta^2 ({\bf x}_{01}
  -{\bf x}_{02}) \ln \left( \frac{x_{01}}{\rho} \right) \right] N_0
  ({\bf x}_{02},{\bf b} + \frac{1}{2} {\bf x}_{12}, Y)
\end{eqnarray*}
\begin{eqnarray}\label{diffn}
  - \frac{\alpha C_F}{\pi^2} \, \int d^2 x_2 \,
  \frac{x^2_{01}}{x^2_{02} x^2_{12}} \, N_0 ({\bf x}_{02},{\bf b} +
  \frac{1}{2} {\bf x}_{12}, Y) \, N_0 ({\bf x}_{12},{\bf b} +
  \frac{1}{2} {\bf x}_{02}, Y). 
\end{eqnarray}
The initial condition for Eq. (\ref{diffn}) is different from that for
Eq. (\ref{fdiff})
\be\label{inn} 
N_0 ({\bf x}_{01},{\bf b}, Y=0) = \gamma ({\bf x}_{01},{\bf b})
\ee
with $\gamma ({\bf x}_{01},{\bf b})$ given by Eq. (\ref{gla}). From
Eq. (\ref{fdef}) one can see now that in order to find $N^D ({\bf
x}_{01},{\bf b}, Y, Y_0)$ one has to solve the same equation
[(\ref{fdiff}) or (\ref{diffn})] for two different initial conditions
given by formulae (\ref{inf}) and (\ref{inn}), which would yield us
with the results for $N_0$ and ${\cal F}$. Then $N^D$ could be found
using Eq. (\ref{fdef}).

As was mentioned above the exact analytical solution of
Eq. (\ref{fdiff}), or, equivalently, (\ref{diffn}) has not been found
\cite{me2,LT}. Nevertheless the qualitative behavior of the solution
of this equation is understood very well, with some quantitative
estimates performed \cite{me2,LT}. It turned out that the solution of
this equation can be qualitatively well approximated by the solution
of the equation with suppressed transverse coordinate dependence
\be \label{toy1}
\frac{\partial N_0 (Y)}{\partial Y} = (\alpha_P - 1) \, N_0 (Y) - 
(\alpha_P - 1) \, N_0 (Y)^2 .
\ee
Eq. (\ref{toy1}) is a toy model of the full Eq. (\ref{diffn}), with
all the transverse coordinate integrations suppressed and the dipole
BFKL kernel substituted by its eigenvalue at the BFKL saddle point.
The fact that the coefficients in front of the linear and quadratic
terms in Eq. (\ref{toy1}) are identical could, for instance, be
understood in the double logarithmic limit ($x_{02}, x_{12} \ll
x_{01}$), in which the kernel in front of the quadratic term in
Eq. (\ref{diffn}) produces an extra factor of two and becomes equal to
the kernel in front of the linear term \cite{me2,LT}. However it is a
more general property of the solution of Eq. (\ref{diffn}), which is
valid not only in the double logarithmic limit.

In this paper we will try to understand the qualitative behavior of
the solution of Eq. (\ref{eqND}) using the toy model of
Eq. (\ref{toy1}). An exact, probably numerical, solution of
Eq. (\ref{diffn}) would give more precise results for $N^D$. However,
we believe that the qualitative features provided by the toy model of
Eq. (\ref{toy1}) will be preserved in the exact solution.

Let us suppose that $N_0 (Y)$ is a solution of Eq. (\ref{diffn}), or,
in our toy model, of Eq. (\ref{toy1}). Then ${\cal F} (Y, Y_0)$ will
be a solution of the same equation with different initial condition
set at a different value of rapidity (see Eq. (\ref{inf})). Thus, if
we neglect the transverse coordinate dependence, the solution for
${\cal F} (Y, Y_0)$ could be obtained from the solution for $N_0 (Y)$
just by a shift in rapidity. Define the shift variable $\Delta Y$ by
\be\label{shift}
N_0 (Y_0 + \Delta Y) = 2 \, N_0 (Y_0) - N_0 (Y_0)^2
\ee
so that 
\be\label{solf}
{\cal F} (Y, Y_0) = N_0 (Y + \Delta Y).
\ee
Now one can see that since $N_0 (Y)$ satisfies the differential
equation (\ref{diffn}) (or Eq. (\ref{toy1}) in our toy model), then
${\cal F} (Y, Y_0)$ given by Eq. (\ref{solf}) satisfies the same
equation (\ref{toy1}) with the initial condition of
Eq. (\ref{inf}). Of course this is true only when one neglects the
transverse coordinate dependence of $N_0$ and ${\cal F}$, but could
also be a reasonable approximation for the case of weak transverse
coordinate dependence.

Solving Eq. (\ref{toy1}) with the initial condition $N_0 (Y=0)=
\gamma$ we obtain
\be\label{notoy}
N_0 (Y) = \frac{\gamma e^{(\alpha_P - 1) Y}}{1 + \gamma (e^{(\alpha_P
- 1) Y} - 1)},
\ee
which employing Eq. (\ref{shift}) yields us with the following
expression for $\Delta Y$
\be \label{ystar}
\Delta Y = \frac{1}{\alpha_P - 1} \ln \left( 2 + \frac{\gamma}{1 -
\gamma} e^{(\alpha_P - 1) Y_0} \right).
\ee
Eq. (\ref{solf}) allows us to write down a toy model expression for
${\cal F}$
\be\label{ftoy}
{\cal F} (Y, Y_0) = \frac{\gamma e^{(\alpha_P - 1) Y} \left( 2 +
\frac{\gamma}{1 - \gamma}  e^{(\alpha_P - 1) Y_0} \right) }{1 - \gamma 
+ \gamma e^{(\alpha_P - 1) Y} \left( 2 + \frac{\gamma}{1 - \gamma}
e^{(\alpha_P - 1) Y_0} \right)}.
\ee
Now we can explore the qualitative behavior of our result. First we
note that for not very large rapidities, $Y \sim 1/\alpha$, we can put
denominators in Eqs. (\ref{notoy}) and (\ref{ftoy}) to be equal to
$1$, which through Eq. (\ref{fdef}) provides us with
\be\label{lond}
- \frac{\partial N^D (Y, Y_0)}{\partial Y_0} \approx (\alpha_P - 1)
\gamma^2 e^{(\alpha_P - 1) (Y+Y_0)} + \delta (Y - Y_0) N_0^2 (Y_0).
\ee
The first term on the right hand side of Eq. (\ref{lond}) is the usual
result of the lowest order approach employing only one triple pomeron
vertex \cite{Mueller2,3p} (see Fig. \ref{3pom}). The second
(delta-function) term in Eq. (\ref{lond}) corresponds to the
contribution of the elastic cross section when the rapidity gap fills
the whole rapidity interval. 

Remember that $N^D (Y, Y_0)$ is the cross section of having a rapidity
gap greater or equal to $Y_0$. If one wants to obtain the cross
section for a fixed size of the rapidity gap one has to differentiate
$N^D$ with respect to $Y_0$ and take a negative of the result, since
$N^D$ is obviously a decreasing function of $Y_0$. This is what was
done in arriving at Eq. (\ref{lond}). That way we have verified that
the toy model solution of our equation (\ref{eqND}) maps onto the old
and well established triple pomeron vertex result \cite{3p}.

At the lowest order $\gamma \sim \alpha^2 A^{1/3}$. Thus the first
term on the right hand side of Eq. (\ref{lond}), which is usually
calculated in the perturbative approaches \cite{3p}, is of the order
of $\alpha^5 \ll 1$. We note that using the nonlinear evolution
equation (\ref{eqN}) we were able to obtain control over the kinematic
regions where the coupling is still small but the cross sections can
become very large, with $N_0 \sim 1$. Thus the traditional
perturbative calculation of Eq. (\ref{lond}) appears like a small
perturbation compared to our large cross section result of
Eq. (\ref{notoy}), which was still obtained in the small coupling
constant limit.

At very high energies, when $Y \rightarrow \infty$, one can see from
Eqs. (\ref{notoy}),(\ref{ftoy}) and (\ref{fdef}) that 
\be
N^D (Y, Y_0) \rightarrow 1, \, \, \,  Y \rightarrow \infty.
\ee
This result is easy to understand. As we know from quantum mechanics
at very high energies, when the total cross section is black, only the
elastic and inelastic contributions to the cross section survive to
give half of the total cross section each. The total cross section
becomes $\sigma_{tot} = 2 \pi R^2$, whereas totally inelastic and
elastic cross sections will be $\sigma_{inel}=\sigma_{el}=\pi
R^2$. All the intermediate contributions with finite size rapidity
gaps covering fraction of the total rapidity interval go to
zero. Since $N^D (Y, Y_0)$ is the cross section of having rapidity gap
greater or equal than $Y_0$ for any finite non-zero $Y_0$ it includes
the contribution of the elastic cross section, which corresponds to
the case when the rapidity gap covers the whole rapidity interval. At
the same time $N^D (Y, Y_0)$ does not include the totally inelastic
contribution, when there is no rapidity gap at all. Thus as at $Y
\rightarrow \infty$ only the elastic piece in $N^D$ survives to give half of 
the total cross section, which corresponds to $N^D \rightarrow 1$ in
our notation, where everything has to be multiplied by $\pi R^2$.

Let us define the cross section of single diffractive dissociation
with the fixed rapidity gap $Y_0$, which, for $Y_0 \le Y$, in the
light of the above discussion is
\be 
R ({\bf x}_{01},{\bf b},Y, Y_0) = - \frac{\partial N^D ({\bf
x}_{01},{\bf b},Y, Y_0)}{\partial Y_0} = \frac{\partial {\cal F} ({\bf
x}_{01},{\bf b},Y, Y_0)}{\partial Y_0} + \delta (Y - Y_0) N_0^2 ({\bf
x}_{01},{\bf b},Y_0).
\ee
In the toy model we have
\be \label{fixg}
R (Y, Y_0) = \frac{\gamma^2 (1 - \gamma)^2 (\alpha_P - 1) e^{(\alpha_P
- 1) (Y + Y_0)}}{\left[ 1 + 2 \gamma \left(e^{(\alpha_P - 1) Y} - 1
\right) + \gamma^2 \left(1 - 2 e^{(\alpha_P - 1) Y} + e^{(\alpha_P -
1) (Y + Y_0)} \right) \right]^2} + \delta (Y - Y_0) N_0^2 (Y_0).
\ee
Now one can see that the triple pomeron vertex result of
Eq. (\ref{lond}) could be recovered from the first term of
Eq. (\ref{fixg}) by putting the denominator to be $1$ and neglecting
$\gamma$ with respect to $1$ in the numerator. Also the quantum
mechanical result discussed above is illustrated explicitly: for any
finite $Y_0$ the cross section of a fixed size gap $R (Y, Y_0)$ goes
to zero as $Y \rightarrow
\infty$, as could be derived from Eq. (\ref{fixg}).

\begin{figure}
\begin{center}
\epsfxsize=12cm
\leavevmode
\hbox{ \epsffile{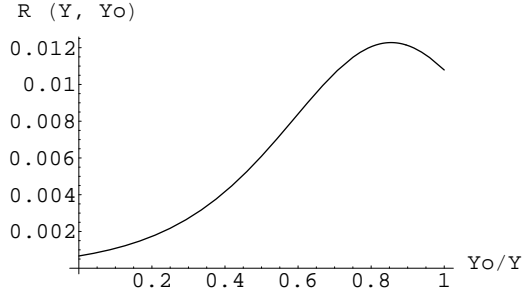}}
\end{center}
\caption{The cross section of diffractive dissociation in units of 
$\pi R^2$ at fixed $Y$ as a function of $Y_0$, which is scaled by
$Y$. The elastic contribution is not included.}
\label{peak}
\end{figure}

Finally, let us plot the cross section of diffractive dissociation $R
(Y, Y_0)$ of Eq. (\ref{fixg}) as a function of the size of rapidity
gap $Y_0$ for a fixed center of mass rapidity $Y$ excluding the
elastic cross section (delta-function) contribution. This is
equivalent to plotting the cross section as a function of the
invariant mass of the particles produced in the virtual photon's
decay, since $Y_0 = Y - \ln M_X^2/Q^2$ (see Fig. \ref{rapgap}). The
plot is shown in Fig. \ref{peak}. There we put $\alpha_P - 1 = .5$,
$\gamma = .03$, $Y=10$ and we divided $Y_0$ by $Y$ on the horizontal
axis.

Fig. \ref{peak} demonstrates that the cross section increases with the
size of the rapidity gap over most of the rapidity interval. This is
what one would expect from the lowest order formula (\ref{lond}). Also
that implies that it is more advantageous to produce particles with
smaller invariant mass $M_X^2$.  However at some very large size of
the rapidity gap the cross section $R (Y, Y_0)$ reaches a maximum and
starts decreasing. The maximum of $R (Y, Y_0)$ in Eq. (\ref{fixg}) is
reached at the size of the rapidity gap
\be \label{yom}
Y_0^{max} = \frac{1}{\alpha_P - 1} \, \ln \left( \frac{2 (1 -
\gamma)}{\gamma} + \frac{(1-\gamma)^2}{\gamma^2} e^{- (\alpha_P - 1)
 Y} \right). 
\ee
From Eq. (\ref{yom}) we can immediately conclude that if the target
nucleus is very large so that each of the dipoles undergoes many
rescatterings making $\gamma \approx 1$ the maximum will be pushed all
the way into the region of negative $Y_0$, making the cross section of
Fig. \ref{peak} just a decreasing function of $Y_0$ for all rapidities
of interest. Thus, if the effects of multiple rescatterings of
individual dipoles become important then the qualitative behavior of
the cross section of diffractive dissociation will change
completely. The effect of multiple rescattering is to push the maximum
of Eq. (\ref{yom}) towards the smaller values of rapidity, which are
easier to reach experimentally.

If the multiple dipole rescatterings are not very important we can
just use the lowest order (one rescattering) $\gamma \sim \alpha^2
A^{1/3} \ll 1$ and consider two limits.  At not very large values of
rapidity, $Y \sim 1/\alpha$, when the single BFKL pomeron exchange is
most important the maximum can never be reached since 
\be
Y_0^{max} \sim
\frac{1}{\alpha_P - 1} \, \ln \frac{1}{\alpha^2 A^{1/3}} \gg \frac{1}
{\alpha_P - 1}.
\ee 
As energy increases and becomes of the order of saturation energy,
corresponding to rapidities of the order of 
\be\label{sat}
Y_{sat} \sim \frac{1}{\alpha_P - 1} \, \ln \frac{1}{\alpha^2 A^{1/3}}
\ee
the maximum will be reached and the effect should be observed.

We can conclude that experimental observation of the maximum of the
cross section of diffractive dissociation at certain size of the
rapidity gap would signify the presence of multiple rescattering
effects either in multiple pomeron exchanges, which are encoded in the
branching of the dipole wave function, or in multiple interactions of
each of the dipoles with the target, which were enclosed in the
initial condition (\ref{gla}). Therefore it would be an independent
test of whether the evolution equations describing the hadronic or
nuclear structure functions should be still used at the linear order
or non-linear effects like in Eq. (\ref{eqN}) should be taken into
account. The qualitative conclusion will also be independent of the
non-perturbative initial conditions of the evolution equations. Here
we have to warn the reader that our results in this section are based
on the toy model and a more elaborate numerical analysis of
Eq. (\ref{eqN}) is needed to achieve the required level of rigor in
these conclusions. Also Eq. (\ref{eqN}) resums fan diagrams, leaving
out the pomeron loop contributions. Even though pomeron loops are
always suppressed by powers of $A$ compared to fan diagrams, they will
become important at the rapidities of the order of $Y \sim
\frac{1}{\alpha_P - 1} \, \ln \frac{1}{\alpha^2}$. This is still above
the saturation energy of Eq. (\ref{sat}) and does not affect much the
kinematic region considered above. Nevertheless for real life nuclei
the suppression of pomeron loop diagrams is not that large and they
may play an important role already at the rapidities of interest.

\section{Conclusions}

In this paper we have derived the small-$x$ evolution equation for the
cross section of single diffractive dissociation (\ref{eqND}). It
resums multiple BFKL pomeron exchange diagrams contributing to the
diffractive cross section. The equation is not derived in the
framework of any model of multiple exchanges: it was directly derived
from QCD. We made use of the fact that in DIS with large $Q^2$ the
strong coupling constant is small. We also employed the leading
logarithmic approximation resumming logarithms of Bjorken $x$
\cite{EAK,Yay}. Finally we employed the large $N_c$ limit of QCD at
high energies --- Mueller's dipole model
\cite{Mueller1,Mueller2,Mueller3,MZ}.

The obtained equation (\ref{eqND}) is nonlinear and most likely can
not be solved analytically. We solved a simplified model of the
equation (Eqs. (\ref{notoy}) and (\ref{ftoy})), in which the
transverse coordinate dependence was suppressed. We believe that this
approximation preserves the qualitative features of the solution. In
this toy model we observed an interesting phenomenon: the diffractive
cross section has a maximum as a function of the rapidity gap $Y_0$
(Fig. \ref{peak}). This effect can not be obtained from the usual
single triple pomeron vertex approach of \cite{3p,Mueller2} and if
experimentally observed would signify the importance of non-linear
effects in QCD evolution of structure functions. The effect should be
seen in the physically measurable quantity --- DIS diffractive cross
section and it would be an independent evidence of the onset of
saturation.

\section*{Acknowledgments}

The authors would like to thank Larry McLerran and Al Mueller for
several helpful and encouraging discussions. We have also benefited
from interesting discussions with Errol Gotsman, Uri Maor, Kirill
Tuchin and Heribert Weigert.

This work was carried out while E.L. was on sabbatical leave at
BNL. E.L. wants to thank the nuclear theory group at BNL for their
hospitality and support during that time.

The research of E.L. was supported in part by the Israel Science
Foundation, founded by the Israeli Academy of Science and Humanities,
and BSF $\#$ 9800276. This manuscript has been authorized under
Contract No. DE-AC02-98CH10886 with the U.S. Department of Energy.


\begin{thebibliography}{99}
  
\bibitem{Mueller1} A.H.\ Mueller, Nucl.\ Phys.\ {\bf B415}, 373
  (1994).

\bibitem{Mueller2} 
A.H.\ Mueller and B. Patel, Nucl.\ Phys.\ {\bf B425}, 471 (1994).

\bibitem{Mueller3}
 A.H.\ Mueller, Nucl.\ Phys.\ {\bf B437}, 107 (1995).

\bibitem{MZ}
 Z.\ Chen, A.H.\ Mueller, Nucl.\ Phys.\ {\bf B451}, 579 (1995).

\bibitem{EAK} 
  E.A. Kuraev, L.N. Lipatov and V.S. Fadin, {\em Sov.
    Phys. JETP} {\bf 45}, 199 (1977).

\bibitem{Yay} 
  Ya.Ya. Balitsky and L.N. Lipatov, {\em Sov. J. Nucl.
    Phys.} {\bf 28}, 22 (1978).

\bibitem{me} 
Yu. V. Kovchegov, Phys. Rev. D {\bf 60}, 034008 (1999).

\bibitem{GLR} 
  L.V.\ Gribov, E.M.\ Levin, and M.G.\ Ryskin, Nucl.\ Phys.\ {\bf
  B188}, 555 (1981); Phys.\ Reports {\bf 100}, 1 (1983).

\bibitem{Bali} I. I. Balitsky, hep-ph/9706411; Nucl. Phys. {\bf B463},
  99 (1996).

\bibitem{Mueller4}
A.H.\ Mueller, Nucl.\ Phys.\ {\bf B335}, 115 (1990).

\bibitem{me2} 
Yu. V. Kovchegov, hep-ph/9905214.

\bibitem{shw} A. Schwimmer, Nucl. Phys. {\bf B 94}, 445 (1975).

\bibitem{AGL} A.L.\ Ayala, M.B.\ Gay Ducati, and E.M.\ Levin, Nucl.\ 
  Phys.\ {\bf B493}, 305 (1997); Nucl.\ Phys.\ {\bf B551}, 355 (1998).

\bibitem{LT} E. Levin and K. Tuchin, hep-ph/9908317.

\bibitem{3p} A. H. Mueller, Phys. Rev. {\bf D2}, 2963 (1970); Phys. Rev. 
{\bf D2}, 150 (1971).

\bibitem{agk}
V. A. Abramovskii, V. N Gribov, and O. V. Kancheli,
Sov. J. Nucl. Phys. , Vol. 8, No. 3, (1974).  

\bibitem{wm}
M Wusthoff and A. D. Martin, hep-ph/9909362; E. Levin and M. Wusthoff,
Phys. Rev. {\bf D 50}, 4306 (1994).

\bibitem{km} 
 Yu. V. Kovchegov, A.H. Mueller, Nucl. Phys. {\bf B 529}, 451 (1998).

\bibitem{mL}
Yu. V. Kovchegov, L. McLerran, Phys. Rev. D {\bf 60}, 054025 (1999). 

\bibitem{mu} A. H. Mueller, Eur. Phys. J. {\bf A1}, 19(1998).

\bibitem{Fro} M. Froissart, Phys. Rev. {\bf 123}, 1053 (1961); E.
    Levin, hep-ph/9808486 and references therein.

\bibitem{br}
J. Bartels, M.G. Ryskin, Z. Phys. {\bf C76},241 (1997) and references
therein.
 
\bibitem{pomdif}
S. Bondarenko, E. Gotsman, E. Levin, and U. Maor, in
preparation. 


\end{thebibliography}
\end{document}